\documentclass[onecolumn,teaser=inline,titlespacing=compact]{mxbai}

\usepackage{colortbl}
\definecolor{avgcol}{HTML}{F4F4F5}  
\colorlet{dealrow}{mxbTint}         

\makeatletter
\newlength\narrowfigwidth
\g@addto@macro\@floatboxreset{\def\@tempa{figure}\ifx\@captype\@tempa
  \setlength\columnwidth{\narrowfigwidth}\fi}
\def\AnonRepoURL{https://anonymous.4open.science/r/dealrank}
\newcommand\RepoURL{https://github.com/mixedbread-ai/dealrank}
\let\mxbp@url\url
\newcommand\mxbp@swapurl[1]{\def\@tempa{#1}\ifx\@tempa\AnonRepoURL
  \expandafter\mxbp@url\expandafter{\RepoURL}\else\mxbp@url{#1}\fi}
\newcommand\inputwithrepo[1]{\begingroup\let\url\mxbp@swapurl\input{#1}\endgroup}
\let\mxbp@input\input
\newcommand\inputmethod{%
  \@namedef{input}##1{\ifstrequal{##1}{tables/main_gemini}{}{\mxbp@input{##1}}}%
  \mxbp@input{sections/method}\let\input\mxbp@input}

\newcommand\mxbp@bottomtable{\let\mxbp@table\table\let\mxbp@endtable\endtable
  \renewenvironment{table}[1][]{\mxbp@table[!b]}{\mxbp@endtable}}
\newcommand\inputresults{%
  \@namedef{input}##1{\ifstrequal{##1}{tables/models}{\mxbp@input{tables/main_gemini}}{}%
    \ifstrequal{##1}{tables/depth}{\begingroup\mxbp@bottomtable\mxbp@input{##1}\endgroup}{\mxbp@input{##1}}}%
  \mxbp@input{sections/results}\let\input\mxbp@input}
\makeatother

\newcommand\RefSep{2pt}

\AtBeginDocument{\setlength\bibsep{\RefSep plus 1pt}%
  }

\title[Deal With It]{Deal With It: On Better Schedules\texorpdfstring{\\}{ }for Listwise Reranking}
\author[1,2]{Benjamin Clavié}[ben@mixedbread.com]   
\author[1]{Herumb Shandilya}
\author[1]{Hanna Lichtenberg}
\author[1]{Aamir Shakir}
\author[2,3]{Makoto P. Kato}                        
\affiliation[1]{Mixedbread}                         
\affiliation[2]{National Institute of Informatics (NII)}
\affiliation[3]{University of Tsukuba}
\reportlink[mixedbread-ai/dealrank]{Code}{https://github.com/mixedbread-ai/dealrank}
\reportlink[Deal With It]{Blog}{https://mixedbread.com/blog/dealrank}
\reporttype{Preprint}
\date{November 2026}

\begin{document}

\begin{abstract}
\setlength\parindent{0pt}
Listwise LLM rerankers cannot reliably rank hundreds of documents in a single call, and rely on schedules to orchestrate their calls. Parallel schedules split candidates into groups and advance a fixed quota from each. When a group holds more of the top passages than its quota, some are cut, and expensive repair calls become required. 
We show that the way candidates are split matters where such a quota can bind, and that, in expected cuts, the best split there deals them round-robin by retriever rank, like cards. Although not conceptually new, dealing has remained unanalyzed. We prove that, under a simple retriever model, no split with equal quotas cuts fewer of the true top ten on average, and that contiguous chunks cut more as the retriever improves. DealRank, the simplest schedule our analysis allows, is on average within 0.1 nDCG@10 points of the best schedules over six LLMs, in two rounds. It is consistently 1.1 to 1.4 times faster than BlitzRank with concurrent first calls up to a thousand candidates. There, it needs three rounds and is eleven times faster than the sliding window.
In existing schedules such as BlitzRank, dealing reduces repair calls from 59\% of queries to 14\%, improving latency with no significant change in quality. Random partitioning appears equivalent at a quota of four, as predicted, but ties results to the seed and trails by 1.3 nDCG@10 points at a quota of one over a dense retriever.
\end{abstract}
\begin{teaserfigure}
  \centering
  \includegraphics[width=\textwidth]{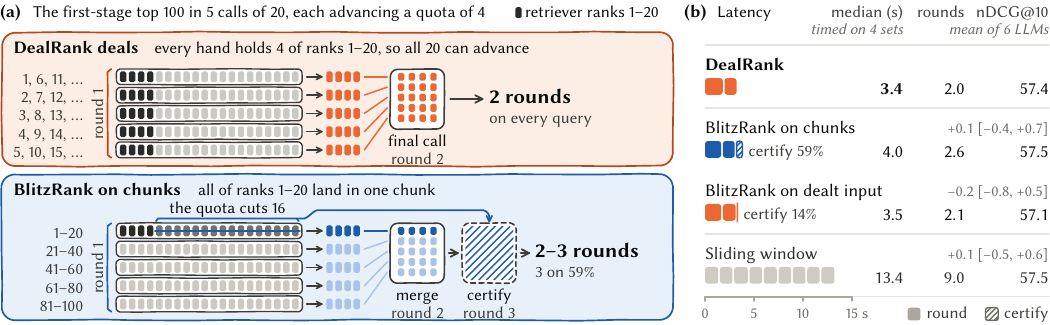}
  \caption{(a)~Dealing against contiguous chunks. (b)~Median Gemini API latency and nDCG@10 gap to DealRank (95\% CI).}
  \Description{Left: two diagrams of 100 retrieved passages split into five calls of twenty, each passage a small card, with the retriever's ranks 1 to 20 drawn dark. In the DealRank diagram hand one holds ranks 1, 6, 11 and so on, so every hand holds four dark cards; each hand advances its four dark cards and the final call holds all twenty, two rounds on every query. In the BlitzRank diagram the calls are ranks 1 to 20, 21 to 40 and so on, so the first call holds all twenty dark cards, of which the four that fit the quota stay and the sixteen it cuts are struck through. The merge call holds four of the twenty, and an arrow runs from the sixteen cut cards to a hatched, dashed certification call, a third round on 59 percent of queries, so two to three rounds. Right: horizontal bars of median latency in seconds, one block per round, with mean rounds and nDCG@10 averaged over six LLMs: DealRank 3.4 seconds in 2.0 rounds, 57.4; BlitzRank on chunks 4.0 seconds in 2.6 rounds with a certification block on 59 percent of queries, 57.5; BlitzRank on dealt input 3.5 seconds in 2.1 rounds with a certification block on 14 percent of queries, 57.1; sliding window 13.4 seconds in 9.0 rounds, 57.5. Every nDCG@10 difference to DealRank has a 95 percent interval that includes zero.}
  \label{fig:hero}
\end{teaserfigure}

\maketitle

\section{Introduction}

Listwise reranking with large language models (LLMs), where the model is provided with a query and a certain number of documents to order according to relevance, has experienced significant interest due to its high quality~\cite{sun2023rankgpt,ma2023lrl,pradeep2023rankvicuna,pradeep2023rankzephyr,gangireddy2024first}. This method is primarily used in the typical ``retrieve-then-rerank'' 2-stage pipeline~\cite{nogueira2019passage}, where a first-stage retriever, commonly BM25~\cite{robertson2009bm25} or neural-based methods such as dense embeddings~\cite{karpukhin2020dpr}, SPLADE sparse embeddings~\cite{formal2021splade} or ColBERT-inspired~\cite{khattab2020colbert} multi-vector embeddings, first retrieves the candidates. The common practice is to rerank a small, but not insignificant number of documents, with 100 and 1000 being commonly evaluated targets~\cite{sun2023rankgpt,craswell2020trecdl}.

Listwise rerankers experience latency and context window constraints, either due to models supporting small windows~\cite{pradeep2023rankzephyr} or models experiencing context rot~\cite{hong2025contextrot}, with degraded performance as window sizes climb. As such, they require a \emph{schedule}, which decides which windows to rank and in what order, in order to produce the final ranking.

The sliding window of the original RankGPT~\cite{sun2023rankgpt}, which covers a hundred candidates from the bottom up in nine calls, each waiting for the one before, is suboptimal. The problem is a known one, as presented by~\citet{blitzrank2026}: given an arbitrary number of horses, how many races do we need to accurately identify the \textbf{\textit{n}} top performers?

Many schedules aiming to fix this have been proposed: setwise approaches~\cite{zhuang2024setwise}, tournament brackets~\cite{chen2025tourrank}, highly-parallelized approaches~\cite{dedov2025jointrank}, among others. \citet{blitzrank2026} correctly identify that each of these methods leaves out substantial information by under-utilizing the implicit pairwise relationships between documents, and seek to address this with a principled tournament approach, \textit{BlitzRank}. This tournament achieves matching ranking quality at significantly reduced token counts.

Token counts, however, say little about how long a user waits. Independent calls can run at the same time, so the wait is set by the chain of \emph{rounds}, where each round is a set of calls that must wait for an earlier round. BlitzRank notes that disjoint calls can run in parallel, but does not further explore parallel depth and its implications~\cite{blitzrank2026}. A parallel schedule splits the candidates into groups, ranks the groups at once and advances a fixed \emph{quota} from each. When one group holds more of the top passages than its quota, good passages are cut, and additional rounds, such as BlitzRank's certification or TourRank's further stages~\cite{chen2025tourrank}, must be added to recover them. In this short paper, we demonstrate that the split matters where a group's quota can bind. We subsequently argue that, given that the nature of re-ranking implies a first-stage retriever, the best split deals the candidates round-robin by retriever rank.

The need for dealing is best shown by the \textbf{certification mechanism} methods such as BlitzRank require: in the BlitzRank implementation, the schedule can trigger an additional sequential call, which re-ranks passages that the chunks cut to check that none belongs in the top ten. This call is triggered on 59\% of queries on average, and is needed because a retriever puts relevant passages near the top, so its pre-certification list often removes highly relevant documents. Indeed, its released implementation ranks five contiguous \emph{chunks} of the retriever's top 100 (ranks 1--20, then 21--40, and so on) and sends the top four of each into one merge call. Dealing the list like a deck of cards avoids most of these cuts. Hand~$i$ receives the retriever's ranks $i, i{+}5, i{+}10, \ldots$ (Figure~\ref{fig:hero}), so no hand holds more than four of the retriever's top twenty and all twenty can advance.

Dealing itself is not new. Indeed, it is a generalization of the equal-gap seeding of knockout tournaments~\cite{karpov2016equalgap,karpov2018seedings}, and it has been suggested that approximate top-$k$ kernels on accelerators interleave their buckets, a form of dealing, because contiguous ones lose recall on correlated inputs~\cite{key2024approxtopk}. Among existing schedules, TourRank deals its pool at every stage, while others select around pivots~\cite{listk2026}, run tournaments of small windows~\cite{yoon2024listt5} or re-rank uncertain passages~\cite{yoon2025acurank}. However, there is little analysis of why the split matters, where it does not, or which later calls a dealt split makes unnecessary, rendering existing approaches computationally wasteful. 

\subsection{Contribution}

In this paper, we address the above questions with the following findings.

First, where a quota can bind, dealing is the optimal split under a simple retriever model. We prove that, under this model, no split into groups with equal quotas cuts fewer of the true top ten on average. Moreover, the only splits as balanced as dealing at every retriever depth are other deals. Random groups appear equivalent to dealing at a quota of four, as the analysis predicts, but are balanced only on average, so they keep cutting however good the retriever is, and their results depend on the seed. Contiguous chunks cut more as the retriever improves, and already cost 2.7 nDCG@10 points over BM25.

Second, most schedules reach similar nDCG@10 but differ widely in calls, rounds, tokens and latency. DealRank, the simplest schedule built on these findings, parallelizes all initially dealt hands at once before adjusting its final rankings in a single-call final round. It reaches performance competitive with all best-performing schedules, and outperforms weaker dealing-based ones such as TourRank with one tournament, with fewer ranking rounds than any other multi-round schedule. It additionally comes with consistently predictable numbers of rounds: the total number of rounds is fixed and based only on the number of input documents, unlike other schedules which may require additional certification calls~\cite{blitzrank2026}.

Third, across five schedules, dealing helps wherever a quota can bind (\S\ref{sec:results}). It removes most of BlitzRank's certification calls without statistically significant changes in its nDCG@10. On the other hand, contiguous groups would cost TourRank, which already deals, 2.2 to 4.5 points. In TDPart, ListK and BlitzRank at $w{=}10$, where no quota can bind, dealing behaves as expected, without any major changes in quality.

Building upon this, we demonstrate that DealRank, in a hierarchical form, scales considerably better than existing schedules as the list of candidate documents grows. On top 1000 reranking tasks, its ranking performance is again within noise of other strong performing schedules, while needing only three rounds and achieving an 11-fold speedup over the sliding window and a 1.26-fold one over BlitzRank with concurrent first calls.

Finally, we release our implementation of dealing and DealRank, as well as a reproduction of all core experiments in this short paper at \url{https://github.com/mixedbread-ai/dealrank}.

\inputmethod
\inputresults

\section*{Acknowledgements}

We thank Rohan Jha for his insights in discussions about this work during his internship at Mixedbread.

\bibliographystyle{plainnat-eprint}
\bibliography{refs}

\end{document}